\documentstyle[12pt,epsf]{article}

\newcommand{\sect}[1]{\setcounter{equation}{0}\section{#1}}

\begin{document}

\title{Charged Nariai Black Holes With a Dilaton}
\author{{\sc Raphael Bousso}\thanks{\it R.Bousso@damtp.cam.ac.uk}
      \\[1 ex] {\it Department of Applied Mathematics and}
      \\ {\it Theoretical Physics}
      \\ {\it University of Cambridge}
      \\ {\it Silver Street, Cambridge CB3 9EW}
       }
\date{DAMTP/R-96/39 \\[1ex] gr-qc/9608053}

\maketitle

\begin{abstract}

  The Reissner-Nordstr\"om-de~Sitter black holes of standard
  Einstein-Max\-well theory with a cosmological constant have no
  analogue in dilatonic theories with a Liouville potential. The only
  exception are the solutions of maximal mass, the Charged Nariai
  solutions. We show that the structure of the solution space of the
  Dilatonic Charged Nariai black holes is quite different from the
  non-dilatonic case.  Its dimensionality depends on the exponential
  coupling constants of the dilaton. We discuss the possibility of
  pair creating such black holes on a suitable background. We find
  conditions for the existence of Charged Nariai solutions in theories
  with general dilaton potentials, and consider specifically a massive
  dilaton.

\end{abstract}

\pagebreak

\sect{Introduction}

This paper is motivated by two questions which have received
much attention in recent years:
the pair creation of black holes on a cosmological background via
instantons~\cite%
{GinPer83,MelMos89,Rom92,ManRos95,HawRos95b,BouHaw95,BouHaw96},
and
the generalisation of cosmological
black hole solutions to dilatonic theories~\cite%
{HorHor93a,MakShi93,GreHar93,PolWil94,PolTwa95a,PolTwa95b,ChaHor95}.

Charged Nariai (CN) spacetimes are metrics which can be written as the
direct topological product of $1+1$-dimensional de~Sitter space with a
round two-sphere. They represent a pair of black holes immersed in
de~Sitter space. They admit a Euclidean section, which is given by the
topological product of two two-spheres (not necessarily of the same
radius). Thus they mediate a pair creation process on the background
of de~Sitter space, if such a background is available in a given
theory. In this paper we investigate CN spacetimes occuring in
dilatonic theories.

If nature is described by string theory, a dilaton field must be
included in the low-energy gravitational action. The solutions and
solution spaces of such theories can differ significantly from those
of standard Einstein gravity.  Dilatonic black holes have proven
useful for the study of black hole entropy, thermodynamics and pair
creation\cite{GibMae88,DowGau94a,DowGau94b,GarGid94,HawHor95}. It is
therefore of interest to look for dilatonic analogues to the
standard black hole solutions.

Here we start from Einstein-Maxwell theory with a cosmological
constant,
\begin{equation}
L = (-g)^{1/2} \left( R - 2\Lambda - F^2 \right),
 \label{eq-nodil-lagrangian}
\end{equation}
where
\begin{equation}
F^2 = F_{\mu\nu} F^{\mu\nu},\ g = \det(g_{\mu\nu}).
\end{equation}
This Lagrangian admits a three-parameter family of static charged
black hole solutions:
\begin{equation}
ds^2 = - U(r)\, dt^2 + U(r)^{-1} dr^2 + r^2 d\Omega_2^2,
  \label{eq-RNdS}
\end{equation}
where
\begin{equation} 
U(r) = 1 - \frac{2 \mu}{r} + \frac{Q^2}{r^2} - \frac{1}{3} \Lambda
r^2
  \label{eq-U}
\end{equation}
and $d\Omega_2^2$ is the metric on the unit two-sphere.
The black holes can be either magnetically charged,
\begin{equation}
F = Q \sin \theta \, d\theta \wedge d\phi,
\end{equation}
or electrically charged,
\begin{equation}
F = \frac{Q}{r^2} dt \wedge dr.
\end{equation}
They are called Reissner-Nordstr\"om-de~Sitter (RNdS) solutions.

The inclusion of a dilaton field will modify the Lagrangian,
Eq.~(\ref{eq-nodil-lagrangian}). We shall work in the so-called
Einstein frame, where the dilaton field $\phi$ couples to the Maxwell
field and to the cosmological term. The Lagrangian then reads:
\begin{eqnarray}
L = (-g)^{1/2} \left( R - 2(\nabla \phi)^2 - 2\Lambda e^{2b\phi}
    - e^{-2a\phi} F^2 \right).
  \label{eq-dil-lagrangian}
\end{eqnarray}
Because it arises naturally from a cosmological constant term in the
string frame, we have chosen a Liouville potential for the dilaton.
For definiteness, we shall stick to this potential through most of
this paper. However, one can easily extend our methods to more general
potentials; this is discussed in the final section.

The Lagrangian, Eq.~(\ref{eq-dil-lagrangian}), is invariant under the
transformation
\begin{equation}
  a \rightarrow -a,\ b \rightarrow -b,\ \phi \rightarrow -\phi.
\end{equation}
We fix a gauge by choosing $a \geq 0$.  The variation of the action
with respect to the metric, Maxwell, and dilaton fields yields the
following equations of motion:
\begin{eqnarray}
R_{\mu\nu} & = & 2 \nabla_\mu \phi \nabla_\nu \phi
  + g_{\mu\nu} \Lambda e^{2b\phi}
  + 2 e^{-2a\phi} \left( F_{\mu\lambda} F_{\nu}^{\ \lambda}
  - \frac{1}{4} g_{\mu\nu} F^2 \right), 
  \label{eq-dil-motion-g} \\
0 & = & \nabla_{\mu} \left( e^{-2a\phi} F^{\mu\nu} \right),
  \label{eq-dil-motion-F} \\
2 \nabla^2 \phi & = & - a e^{-2a\phi} F^2 + 2 \Lambda b e^{2b\phi}. 
  \label{eq-dil-motion-phi}
\end{eqnarray}

One might hope that these equations give rise to a three-parameter
family of dilaton black holes, which would be analogous to the RNdS
black holes.  Poletti and Wiltshire have shown, however, that such
solutions do not exist for a Liouville potential with a positive
cosmological constant~\cite{PolWil94}. This no-go theorem is related
to the simple fact there exists no static de~Sitter-type solution that
could act as a background. Correspondingly, the theorem only bans
dilatonic RNdS solutions which approach de~Sitter space
asymptotically.

The paper is outlined as follows.  We begin by reviewing some
properties of the RNdS solutions, in Sec.~\ref{sec-nodil}. We point
out that they are all asymptotically de~Sitter except for a
two-dimensional subspace, the CN solutions, which are special in two
ways. Firstly, unlike most other RNdS solutions, they admit smooth
Euclidean sections. Therefore they can be pair created, as we discuss
in Sec.~\ref{sec-nodil-euc}.  Secondly, because of their unusual
asymptotics, the CN solutions are the only kind of RNdS solutions that
avoid the Poletti-Wiltshire no-go theorem.  Thus they can be
generalised to dilatonic theories with a Liouville potential.

The Dilatonic Charged Nariai (DCN) solutions are easily found, since
the dilaton will be constant. Nevertheless, the solution space that
they form turns out to be quite interesting. In Sec.~\ref{sec-DCN} we
show that the range of some parameters differs significantly from the
non-dilatonic case. We find that both the type of black hole charge,
and the dimensionality of the solution space depends on the
exponential coupling constants $a$ and $b$. Like the CN solutions, the
DCN solutions admit a smooth Euclidean section. In
Sec.~\ref{sec-dil-euc} we calculate the Euclidean action of this
instanton, and discuss whether it can mediate black hole pair
creation. Finally, we show in Sec.~\ref{sec-general} how to find DCN
black holes for a general dilaton potential. As an example, we
consider the case of a massive dilaton.

\sect{Standard Reissner-Nordstr\"om-de~Sitter}
  \label{sec-nodil}

\subsection{Solution Space}
\label{sec-nodil-lor}

In the RNdS solutions, Eq.~(\ref{eq-U}) contains three parameters,
$Q$, $\mu$ and $\Lambda$. $Q$ is the charge of the black hole, and
$\mu$ is related to the mass. Strictly speaking, the cosmological
constant $\Lambda$ is not a parameter of the solution space, but
rather of the space of theories.  Since the cosmological constant is
nearly zero in our universe, however, a cosmological term in the
action will realistically be due to the vacuum energy of a scalar
field. In inflationary cosmology, this field starts out large and
changes slowly over time.  Its potential will act like an effective
cosmological constant $\Lambda_{\rm eff} > 0$. For many purposes it
can be approximated by a fixed $\Lambda$, as we do here.
$\Lambda_{\rm eff}$, however, will depend on the value of the inflaton
field. Thus it is a degree of freedom in the solutions, and is not
specified by the theory. It is with a view to applications in
cosmology, therefore, that we include $\Lambda$ among the solution
space parameters.

\begin{figure}[htb]
  \hspace*{\fill} \vbox{\epsfxsize=.6\textwidth \epsfbox{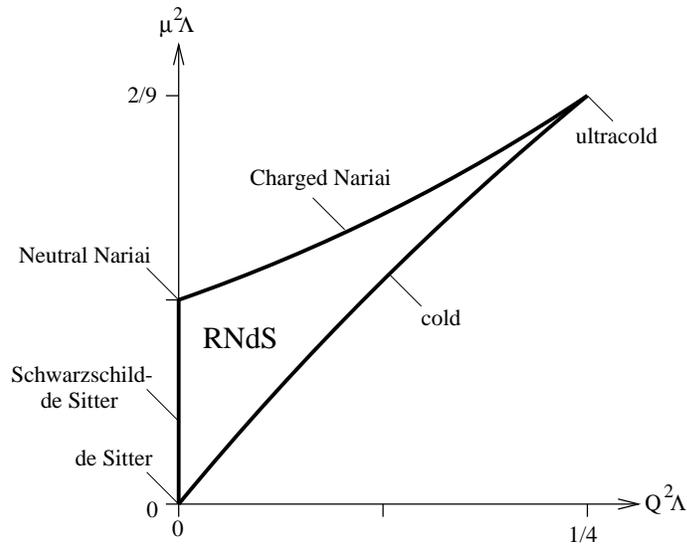}}
  \hspace*{\fill}
\caption[Parameter space of the Reissner-Nordstr\"om-de~Sitter
 solutions]%
 {\small\sl The Reissner-Nordstr\"om-de~Sitter solutions are the
   points on or within the thick line. The plot is of the
   dimensionless quantities $\mu^2 \Lambda$ vs. $Q^2 \Lambda$.}
   \label{fig-RNdS}
\end{figure}

What are the ranges of the three parameters? We shall take $\Lambda$
to be positive. $\mu$ must be non-negative to avoid
naked singularities.
For the same reason, the black hole may not be larger than the
cosmological horizon. This limits its mass:
\begin{equation}
\mu^2 \Lambda \leq \frac{1}{18} \left[ 1 + 12 Q^2 \Lambda + (1 - 4 Q^2
    \Lambda)^{3/2} \right].
\end{equation}
In particular, $\mu^2 \Lambda$ can never exceed $2/9$.  The charge is
limited by the Bogomol'nyi bound, which we give here only
approximately (see~\cite{Rom92} for more details):
\begin{equation}
Q^2 \leq \mu^2 \left[ 1 + \frac{1}{3} (\mu^2 \Lambda) + O(\mu^4
  \Lambda^2) \right].
\end{equation}
Fig.~\ref{fig-RNdS} shows a two-dimensional projection of the
parameter space for a fixed value of $\Lambda$.  As special cases, the
RNdS solutions include de~Sitter space ($Q=\mu=0$) and the
Schwarzschild-de~Sitter solutions ($Q=0$,
$\mu>0$).

\subsection{Asymptotic Structure and Charged Nariai}
  \label{sec-nodil-asy}

The asymptotic structure of the RNdS solutions plays an important and
problematic role when one tries to include a dilaton; therefore we
shall dwell on it for a while.  In a generic region of the parameter
space, $ U $ has three positive roots, which are denoted, in ascending
order, by $ r_{\rm i} $, $ r_{\rm o} $ and $ r_{\rm c} $.  They are
the radii of the inner and outer black hole horizons and the
cosmological horizon.  The causal structure of the spacetime can be
seen from its Carter-Penrose diagram,
\begin{figure}[htb]
  \hspace*{\fill} \vbox{\epsfxsize=.6\textwidth \epsfbox{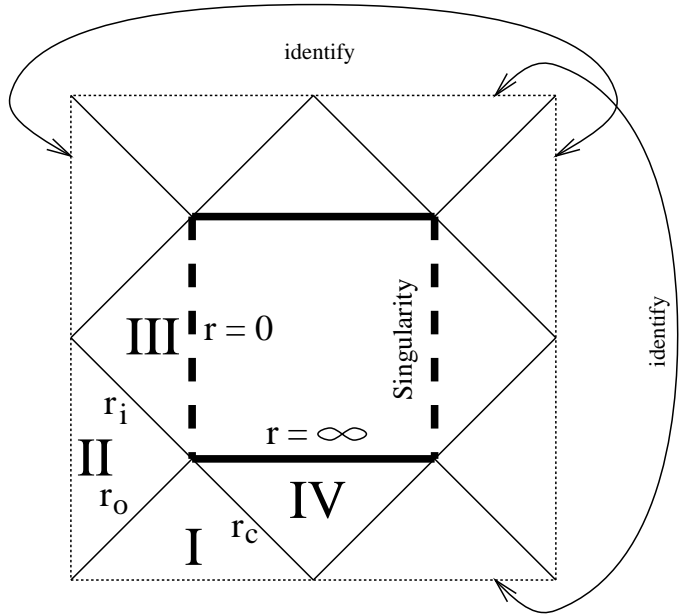}}
  \hspace*{\fill}
\caption[Carter-Penrose diagram for the Reissner-Nordstr\"om-de~Sitter
 solutions]%
 {\small\sl Carter-Penrose diagram for the
   Reissner-Nordstr\"om-de~Sitter solutions}
   \label{fig-CP-RNdS}
\end{figure}
Fig.~\ref{fig-CP-RNdS}.  An observer who ventures outside $r = r_{\rm
  c}$ will find herself in region IV, where $r$ is a time coordinate
and increases without bound.  By Eq.~(\ref{eq-U}), the spacetime looks
more and more like de~Sitter space as $r \rightarrow \infty$:
\begin{equation}
U(r) \rightarrow 1 - \frac{1}{3} \Lambda r^2.
\end{equation}
Therefore the RNdS solutions represent a pair of black holes immersed
in an asymptotically de~Sitter universe. (The fact that it is a pair
can be seen from the Carter-Penrose diagram.)

The causal structure will be different for solutions on the border of
the parameter space, such as the de~Sitter and neutral
Schwarzschild-de~Sitter universes, and the extreme (``cold'') black
holes (which have maximal charge at a given mass). Still, all these
solutions have the same {\em asymptotic} structure, that is, they all
look like de~Sitter space at large distances from the black hole.
Their Carter-Penrose diagrams look different from
Fig.~\ref{fig-CP-RNdS}, but they all contain a region of type IV.

The only exception to this rule are the so-called Charged Nariai (CN)
solutions, the two parameter family of solutions for which the mass is
maximal at a given charge, as indicated in Fig.~\ref{fig-RNdS}. In
these solutions $r_{\rm o} = r_{\rm c}$; therefore $r$ is not a
suitable coordinate for the region between the outer and cosmological
horizons. An appropriate coordinate transformation was first given for
the neutral case in Ref.~\cite{GinPer83}. It shows that as $r_{\rm o}
\rightarrow r_{\rm c}$, the region between $r_{\rm o}$ and $r_{\rm c}$
does not vanish.  A refinement and a more detailed discussion can be
found in the Appendix of Ref.~\cite{BouHaw96}. The transformation can
be readily generalised to include charged solutions~\cite{HawRos95b}.
We set $r_{\rm o} = \rho - \epsilon$, $r_{\rm c} = \rho + \epsilon$.
With the coordinate transformation
\begin{equation}
r = \rho + \epsilon \cos\chi,\ t = \frac{1}{U(\rho)} \epsilon \psi_I,
\end{equation}
the RNdS metric, Eq.~(\ref{eq-RNdS}), becomes
\begin{equation}
ds^2 = \frac{1}{A} \left( - \sin^2\!\chi d\psi_I^2 + d\chi^2 \right)
     + \frac{1}{B} d\Omega_2^2
  \label{eq-CN-metric}
\end{equation}
in the limit $\epsilon \rightarrow 0$, where
\begin{equation}
A = \lim_{\epsilon \rightarrow 0} \frac{U(\rho)}{\epsilon^2},\
B = \frac{1}{\rho^2},
\end{equation}
and $d\Omega_2^2$ is the metric on a round two-sphere of unit radius.
The Maxwell field is given by
\begin{equation}
F = Q \sin \theta \, d\theta \wedge d\phi
  \label{eq-F-mag}
\end{equation}
(and therefore $F^2 = 2 B^2 Q^2$) in the magnetic case, and by
\begin{equation}
F = Q \frac{B}{A} \sin\chi \, d\chi \wedge d\psi_I
  \label{eq-F-el}
\end{equation}
(and therefore $F^2 = - 2 B^2 Q^2$) in the electric case.

It will be useful to define
\begin{equation}
g = \frac{F^2}{2\Lambda}.
\end{equation}
$g$ is positive (negative) for magnetic (electric) black holes.  One
can conveniently express the parameters $A$ and $B$ as
\begin{equation}
A = \Lambda \left( 1 - |g| \right),\
B = \Lambda \left( 1 + |g| \right).
 \label{eq-g-Lambda-to-AB}
\end{equation}
Therefore we have $A<B$ except in the neutral case, when $A=B$. A
metric with $A>B$ does not admit a real Maxwell field. Since $A$ must
be positive, solutions exist only for $|g|<1$. They can be
parametrised by $(\Lambda,g)$ or $(\Lambda, F^2)$, for instance.
Specification of the metric, i.e.\ of $(A,B)$, determines the Maxwell
field up to the type of charge. The same holds for the parameters
$(\Lambda,Q)$.

The CN metric, Eq.~(\ref{eq-CN-metric}), is just the topological
product of $1+1$ dimensional de~Sitter space with a round two-sphere.
It is a homogeneous space-time with the same causal structure as
$1+1$ dimensional de~Sitter space (see
\begin{figure}[htb]
  \hspace*{\fill} \vbox{\epsfxsize=.45\textwidth
  \epsfbox{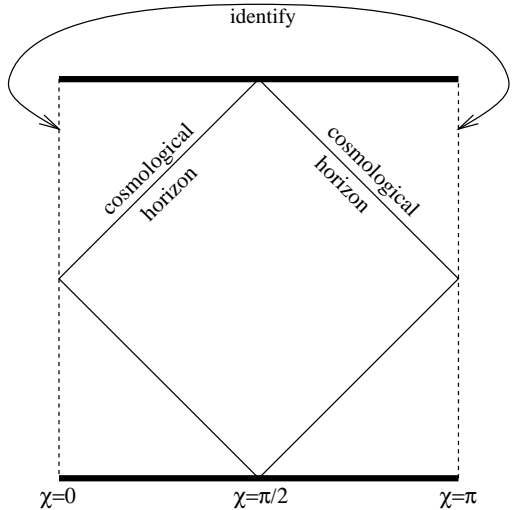}}
  \hspace*{\fill}
\caption[Carter-Penrose diagram for the Charged Nariai solutions]%
{\small\sl Carter-Penrose diagram for the Charged Nariai solutions}
\label{fig-CP-Nariai}
\end{figure}
Fig.~\ref{fig-CP-Nariai}). In this degenerate solution, an observer
sees a cosmological horizon on either side of her (that is, in both
the positive and negative $\chi$ direction). There is no black hole
she could fall into. Like in de~Sitter space, if she crosses the
cosmological horizon (of a different observer), the space-time will
look exactly the same to her as before.

This means, in particular, that the CN solutions do not possess a
de~Sitter-like asymptotic region. The radius of the two-spheres is
constant: $r = B^{-1/2}$ everywhere. Precisely for this reason,
the CN solutions avoid the Poletti-Wiltshire no-go theorem. We shall
show in Sec.~\ref{sec-DCN} that they possess dilatonic analogues.

We should stress that the CN solutions are quantum mechanically
unstable. Because of quantum fluctuations, the radius $B$ of the
two-spheres will vary slightly along the spatial variable $\chi$.
There will be a minimum and a maximum two-sphere, which correspond to
a black hole and a cosmological horizon~\cite{GinPer83,BouHaw96}. Thus
the CN solutions necessarily become ordinary, slightly non-degenerate
RNdS space-times. While this requires only a minute fluctuation of
the horizon radii $r_{\rm o}$ and $r_{\rm c}$, the causal structure
immediately reverts to that of Fig.~\ref{fig-CP-RNdS}.

\subsection{Euclidean Solutions}
  \label{sec-nodil-euc}

Euclidean solutions of the Einstein equations (instantons) can be used
for the semi-classical description of non-perturbative gravitational
processes, such as the spontaneous quantum mechanical creation of a
pair of black holes on some background. Here we are interested in the
pair creation rate of RNdS black holes. The appropriate background is
de~Sitter space, which the black hole solutions approach
asymptotically. (de~Sitter space is also the appropriate background
for the CN solutions, since the quantum fluctuations of the geometry
cannot be neglected in the semi-classical treatment.)  One must find
an instanton which can be analytically continued to match a spacelike
surface $\Sigma$ of the Lorentzian background spacetime, and similarly
one must find an instanton for the spacetime with the black holes.
Then one calculates the Euclidean actions of these solutions, $I_{\rm
  bg}$ and $I_{\rm bh}$. Neglecting a prefactor, the pair creation
rate $\Gamma$ is given by
\begin{equation}
\Gamma =
\exp \left[ - 2 \left( I_{\rm bh} - I_{\rm bg} \right) \right].
\label{eq-pcr}
\end{equation}

Only those Lorentzian black hole solutions which possess regular
Euclidean sections can be pair created. In particular, not all RNdS
solutions can be obtained as analytic continuations of smooth
Euclidean solutions~\cite{MelMos89,Rom92,ManRos95}. Of the entire
three-parameter set, there are only three (intersecting) two-parameter
subsets of RNdS solutions that can be pair created: the cold, the
``lukewarm'' ($Q=\mu$), and the CN black holes. We shall not bother
ourselves with the cold and lukewarm cases, since they cannot be
retained when we introduce a dilaton later. The CN black holes,
however, will have dilatonic analogues, and luckily, they have a
regular Euclidean section:
\begin{equation}
ds^2 = \frac{1}{A} \left( \sin^2\!\chi d\psi_R^2 + d\chi^2 \right)
     + \frac{1}{B} d\Omega_2^2.
  \label{eq-Eu-CN-metric}
\end{equation}
This is the topological product of two round two-spheres of radii
$A^{-1/2}$ and $B^{-1/2}$.  The Maxwell field is given by
\begin{equation}
F = Q \sin \theta \, d\theta \wedge d\phi
\end{equation}
in the magnetic case, and by
\begin{equation}
F = -i Q \frac{B}{A} \sin\chi \, d\chi \wedge d\psi_R
  \label{eq-Eu-F-el}
\end{equation}
in the electric case. The Lorentzian solutions can be recovered by
setting $\psi_R = i \psi_I$.

The Euclidean action is given by
\begin{equation}
I = -\frac{1}{16\pi} \int \! d^4\!x\, g^{1/2}
      \left( R - 2\Lambda - F_{\mu\nu} F^{\mu\nu} \right)
      -\frac{1}{8\pi}  \int_{\Sigma} d^3\!x\, h^{1/2} K,
  \label{eq-nodil-Eu-action}
\end{equation}
where $K$ is the trace of the second fundamental form and $h$ is the
determinant of the three-metric on $\Sigma$.  Since the sign of $F^2$
depends on the type of charge, it would seem that electric and
magnetic black holes would have different pair creation rates. Even
worse, highly charged electric black holes ($g<-1/3$) would have a
lower action than de~Sitter space and their pair creation would thus
be enhanced relative to the background.  However, this problem only
appears because the above action is inappropriate for electrically
charged black holes.  Variation of this action will give the correct
equations of motion only if the gauge potential $A_\nu$ is held fixed
on the boundary $\Sigma$.  This is appropriate for magnetic solutions,
since the gauge potential determines the magnetic charge. Generally,
the charge must be held fixed on variation of the action. In the
electric case, however, this corresponds to fixing $F^{\mu\nu}
n_{\mu}$, rather than the gauge potential, on $\Sigma$.  In order to
retain a stationary action, one must compensate this by introducing an
extra boundary term~\cite{HawRos95b}:
\begin{equation}
I_{\rm(electric)} = I - \frac{1}{4\pi} \int_{\Sigma}
d^3\!x\, h^{1/2} F^{\mu\nu} n_\mu A_\nu,
\end{equation}
where $n_\mu$ is the normal to $\Sigma$.

For the background de~Sitter instanton, the Euclidean action is
\begin{equation}
I_{\rm dS} = - \frac{3\pi}{2\Lambda}.
\end{equation}
For the CN instanton, we obtain an action of $- \pi/\Lambda(1+g)$ in
the magnetic case, and $- \pi/\Lambda(1-g)$ in the electric
case. Thus, for both cases,
\begin{equation}
I_{\rm CN} = - \frac{\pi}{\Lambda (1+|g|)}.
  \label{eq-action-CN}
\end{equation}
As one would hope, electric and magnetic black holes of the same
charge thus have the same action, and the same pair creation rate,
\begin{equation}
\Gamma_{\rm CN} =
\exp \left[ - 2 \left( I_{\rm CN} - I_{\rm dS} \right) \right]
= \exp \left[ - \frac{\pi}{\Lambda} \: \frac{1+3|g|}{1+|g|} \right].
\label{eq-pcr-CN}
\end{equation}

\sect{Dilatonic Charged Nariai Solutions} \label{sec-DCN}

\subsection{Solutions}  \label{sec-DCN-solutions}

We shall now present the dilatonic equivalent to the CN black hole
solutions of the previous section. In a solution that is static and
spherically symmetric, the dilaton can only depend on the radius $r$,
which is fixed in the CN case.  Thus the dilaton will be constant. By
comparing the Lagrangians, Eqs.\ (\ref{eq-nodil-lagrangian}) and
(\ref{eq-dil-lagrangian}), we see that the introduction of a constant
dilaton leads only to a rescaling of the cosmological constant and the
Maxwell field. For convenience we define
\begin{eqnarray}
\tilde{F}^{\mu\nu} & = & e^{-a\phi} F^{\mu\nu},
  \label{eq-def-Ftilde}
\\
\tilde{\Lambda}    & = & e^{2b\phi} \Lambda,
  \label{eq-def-Lambdatilde}
\\
\tilde{g}          & = & \frac{\tilde{F}^2}{2 \tilde{\Lambda}}.
  \label{eq-def-gtilde}
\end{eqnarray}
With these definitions the solutions take the same form as in the
previous section, with $(\Lambda,g)$ replaced by
$(\tilde{\Lambda},\tilde{g})$:
\begin{equation}
  ds^2 = \frac{1}{\tilde{A}} \left( - \sin^2\!\chi d\psi_I^2 + d\chi^2
\right) + \frac{1}{\tilde{B}} d\Omega_2^2,
  \label{eq-dilCN-metric}
\end{equation}
where
\begin{equation}
\tilde{A} = \tilde{\Lambda} \left( 1 - |\tilde{g}| \right),\
\tilde{B} = \tilde{\Lambda} \left( 1 + |\tilde{g}| \right).
 \label{eq-tildes-to-AB}
\end{equation}
As before, the Maxwell field is given by
\begin{equation}
F = Q \sin \theta \, d\theta \wedge d\phi
  \label{eq-dil-F-mag}
\end{equation}
in the magnetic case ($F^2 = 2 Q^2 \tilde{B}^2$), and by
\begin{equation}
F = Q \frac{B}{A} \sin\chi \, d\chi \wedge d\psi_I
  \label{eq-dil-F-el}
\end{equation}
in the electric case ($F^2 = - 2 Q^2 \tilde{B}^2$).  The value of
$\phi$ can be obtained from the dilaton equation of motion. Since the
dilaton must be constant, the left hand side of
Eq.~(\ref{eq-dil-motion-phi}) vanishes, and we get
\begin{equation}
\phi = \frac{1}{2(a+b)}
       \ln \left( \frac{a}{b} \frac{F^2}{2\Lambda} \right)
     = \frac{1}{2(a-b)}
       \ln \left[ \pm \frac{a}{b} Q^2 \Lambda
       \left(1+|\tilde{g}| \right)^2 \right],
\label{eq-dil-1}
\end{equation}
where the upper (lower) sign is for magnetic (electric) black holes.

Like the CN solutions, these solutions are quantum mechanically
unstable.  Quantum perturbations will cause the separation of the
cosmological and black hole horizons. Since no non-degenerate static
dilatonic RNdS solutions exist, the dilaton will be time dependent
beyond the cosmological horizon in these perturbed solutions.

We shall see below that the space of DCN solutions is quite different
from its non-dilatonic counterpart, in spite of the fact that they can
be written in a deceptively similar form.

\subsection{Solution Space} \label{sec-solution-space}

The CN and DCN solutions take the same form when written in terms of
$(\Lambda,g)$ and $(\tilde{\Lambda},\tilde{g})$. The variables $g$ and
$\tilde{g}$, however, are of a fundamentally different nature. $g$, as
we have seen, parametrises the CN solutions along with
$\Lambda$. ${\tilde{g}}$, on the other hand, is not a parameter of the
solution space, since by Eqs.~(\ref{eq-dil-motion-phi}) and
(\ref{eq-def-gtilde}) $\tilde{g}$ simplifies to
\begin{equation}
\tilde{g} = \frac{b}{a}.
  \label{eq-ab-to-g}
\end{equation}
This simple result is central to the differences between the dilatonic
and non-dilatonic CN solution spaces.  It means
that ${\tilde{g}}$ is completely fixed by the couplings of the theory
and does not constitute a degree of freedom. While most results for
the DCN solutions can be obtained from analogous results for the CN
solutions by replacing $(\Lambda,g)$ with
$(\tilde{\Lambda},\tilde{g})$ throughout, it is important to keep in
mind that ${\tilde{g}}$ is no longer a free parameter; it can be
replaced by $b/a$. In particular, the constraints on $g$ ($|g| < 1$)
will translate into constraints on the dilatonic theories that allow
DCN solutions; this will be discussed in Sec.~\ref{sec-theory-space}
below. Before that, it will be necessary to determine the degrees of
freedom of the DCN solution space, and their range.

For this purpose we shall repeatedly exploit Eq.~(\ref{eq-ab-to-g}).
Firstly, by Eq.~(\ref{eq-ab-to-g}), the type of the black hole charge
is determined by the theory:
\begin{equation}
{\rm sgn}(F^2) = {\rm sgn}(\tilde{g}) = {\rm sgn}(b).
\end{equation}
If $ b $ is positive, the charges are magnetic; if it is negative,
they are electric. Electric-magnetic duality, in this case, is not a
duality between pairs of solutions to the same theory, but between
solutions in two different theories, $b$ and $-b$. We can now simplify
Eq.~(\ref{eq-dil-1}) to
\begin{equation}
\phi  = \frac{1}{2a(1-\tilde{g})}
       \ln \left[ \frac{\left(1+|\tilde{g}| \right)^2}{|\tilde{g}|}
       Q^2 \Lambda \right],
\label{eq-dil-2}
\end{equation}

Secondly, by Eq.~(\ref{eq-ab-to-g}), the DCN metric,
Eq.~(\ref{eq-dilCN-metric}), contains only one degree of freedom,
$\tilde{\Lambda}$, which determines its overall scale. The ratio
$\tilde{A}/\tilde{B}$, which is the more interesting geometrical
information, is fixed from the start by the absolute value of
$\tilde{g}$:
\begin{equation}
\frac{\tilde{A}}{\tilde{B}} = \frac{1-|\tilde{g}|}{1+|\tilde{g}|}
            = \frac{|a - b|}{|a + b|}.
  \label{eq-ab-to-AB}
\end{equation}
 
Where did the other degree of freedom go? The CN metrics of standard
Einstein-Maxwell theory carry two degrees of freedom. A variation of
the field necessarily changes the metric.  This is no longer true in
the dilatonic case. Holding ${\tilde{\Lambda}}$ fixed, one can still
vary $Q$, or equivalently, $\phi$. Let us therefore choose $Q$ as a
second parameter of the DCN solution space. But if the DCN metric does
not contain information about the charge, maybe $Q$ is just a gauge
degree of freedom? The question is whether and how $Q$ can be
measured.

An observer in a CN universe can determine $Q$ by observing the motion
of a single test particle of unit charge.  Sadly, this method fails
for the DCN solutions.  The particle will feel a force of
$|e^{-2a\phi} F^2|^{1/2}$, which is equal to $|\tilde{F}^2|^{1/2}$ and
therefore independent of $Q$ for fixed ${\tilde{\Lambda}}$. It cannot
distinguish between different values of $Q$. The observer is thus
forced to resort to two test particles, each of unit charge. From
their interaction strength she can infer the value of the dilaton.
This in turn determines $Q$.  Therefore $Q$ is a proper degree of
freedom, in spite of the fact that it is not reflected in the metric.

What are the ranges of the parameters of the solution space,
$\tilde{\Lambda}$ and $Q$? Since $\Lambda>0$, ${\tilde{\Lambda}}$ can
take on any positive value by Eq.~(\ref{eq-def-Lambdatilde}).  Freed
from the geometrical constraints that would limit its range in the CN
case, the charge can take on any real value except zero. (The case
$Q=0$, which implies $b=0$, will be discussed below.)

\subsection{Theory Space} \label{sec-theory-space}

The exponential coupling constants $a$ and $b$ form a two-parameter
space of theories: $a, b \in {\rm\bf R}; a \geq 0$.  Let ${\cal
  S}(a,b)$ denote the space of DCN solutions admitted by the theory
$(a,b)$.  We shall show that ${\cal S}(a,b)$ is either
one-dimensional, or two-dimensional, or empty, depending on $a$ and
$b$ (see
\begin{figure}[htb]
  \hspace*{\fill} \vbox{\epsfxsize=.8\textwidth
  \epsfbox{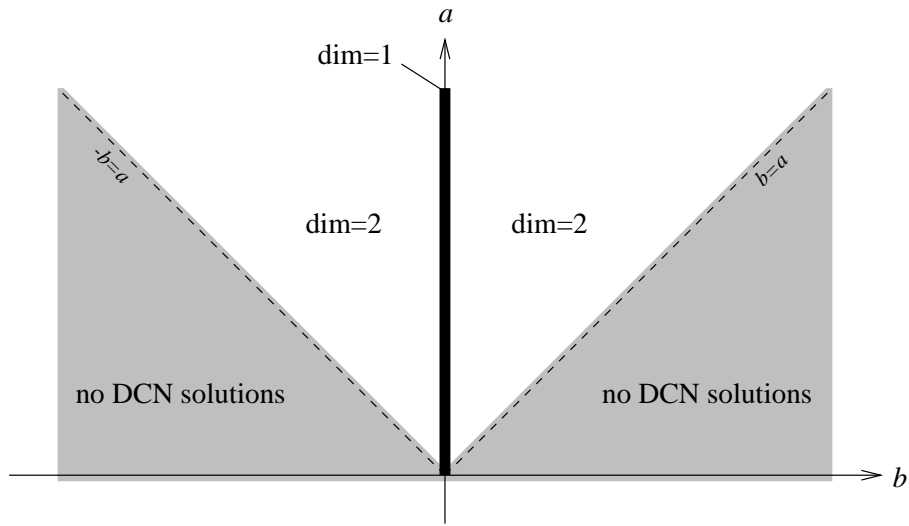}} \hspace*{\fill}
\caption[Classification of dilatonic theory space]%
{\small\sl The dilatonic theories we consider can be labelled by the
  exponential coupling constants $a$ and $b$ that appear in the
  action. For $0<|b|<a$, they admit the two-dimensional DCN solution
  space discussed above. For $b=0$, only a one-dimensional solution
  space is allowed, the non-dilatonic neutral Nariai solutions. For
  other theories, no DCN-type solutions can be found.}
\label{fig-subsets}
\end{figure}
Fig.~\ref{fig-subsets}).

First consider the case $b=0$. By the following chain of equivalences:
\begin{equation}
b=0
\stackrel{(\ref{eq-ab-to-g})}{\Longleftrightarrow}
\tilde{g} = 0
\stackrel{(\ref{eq-def-gtilde})}{\Longleftrightarrow}
\tilde{F}^2 = 0
\stackrel{(\ref{eq-def-Ftilde})}{\Longleftrightarrow}
F^2 = 0
\stackrel{(\ref{eq-dil-F-mag}),\,
  (\ref{eq-dil-F-el})}{\Longleftrightarrow}
Q = 0,
\end{equation}
$b=0$ if and only if $Q=0$. There is no Maxwell field in this case,
and so the dilaton effectively vanishes from the scene.  The
corresponding solution is the non-dilatonic neutral Nariai solution.
$\Lambda = A = B$ is the only degree of freedom.  This is the only
overlap of solution spaces of dilatonic and non-dilatonic theories.

Next consider the solutions described in Sec.~\ref{sec-DCN-solutions}
above, with two degrees of freedom, ${\tilde{\Lambda}}$ and $Q$. For
$b>0$ they are magnetically charged, and for $b<0$ they are
electrically charged. Since $\tilde{A}$ must be positive, we obtain
$|b/a| < 1$ by Eq.~(\ref{eq-tildes-to-AB}). Together with the
condition that $b \neq 0$ and our gauge choice $a \geq 0$, this yields
the range
\begin{equation} 
0 < \frac{|b|}{a} < 1,
  \label{eq-t2}
\end{equation}
for which a two-dimensional space of properly dilatonic CN solutions
is admitted.  It is interesting to note that these geometric
considerations, which fail to limit the charge $Q$ of dilatonic black
holes, constrain instead the range of dilatonic theories which admit
DCN solutions at all.

The remaining theories have $|b| \geq a$, and $(a,b) \neq (0,0)$. By
Eq.~(\ref{eq-ab-to-AB}) this would imply that $\tilde{A}$ were zero or
negative. Therefore no DCN solutions are admitted in this case.

\sect{Euclidean Action and Pair Creation}
  \label{sec-dil-euc}

Just like the non-dilatonic CN solutions of Sec.~\ref{sec-nodil}, the
DCN solutions are analytic continuations of regular Euclidean
solutions, given by the topological product of two round two-spheres:
\begin{equation}
  ds^2 = \frac{1}{\tilde{A}} \left( \sin^2\!\chi d\psi_R^2 + d\chi^2
\right) + \frac{1}{\tilde{B}} d\Omega_2^2,
\end{equation}
with $\tilde{A}$ and $\tilde{B}$ given by Eq.~(\ref{eq-tildes-to-AB}).
The Euclidean dilatonic action is
\begin{eqnarray}
I & = & -\frac{1}{16\pi} \int \! d^4\!x\, g^{1/2}
        \left( R - 2(\nabla \phi)^2 - 2\Lambda e^{2b\phi}
        - e^{-2a\phi} F^2 \right) \nonumber \\
  &   & -\frac{1}{8\pi}  \int_{\Sigma} d^3\!x\, h^{1/2} K.
\end{eqnarray}
The action for the DCN instantons can be obtained from the CN action,
Eq.~(\ref{eq-action-CN}), by replacing $(\Lambda,g)$ with
$(\tilde{\Lambda},\tilde{g})$:
\begin{equation}
I_{\rm DCN} = - \frac {\pi}{\tilde{\Lambda} (1+|\tilde{g}|)}.
  \label{eq-action-DCN}
\end{equation}

These instantons do not, however, readily correspond to a pair
creation process. For that purpose one would need a suitable
background, and we have seen earlier that static de~Sitter-like
solutions do not exist in dilatonic theories. Even the abandonment of
staticity alone will not help, since the dilaton would be pushed
towards negative infinity by the cosmological constant. However, if we
replace the fixed cosmological constant by a slowly decreasing
effective cosmological constant $\Lambda_{\rm eff}$, the dilaton will
only decrease as long as $\Lambda_{\rm eff}$ is large. For $|b| \ll 1$
the dilaton will change slowly over time. Such solutions are well
known in inflationary cosmology. They have been used by a number of
authors~\cite{Lin90,GarLin94,GarWan95} for the description of extended
chaotic inflation.  Because of its similarity to de~Sitter space, an
inflating universe is a suitable background for the pair creation of
black holes~\cite{BouHaw95,BouHaw96}. With the modification $\Lambda
\rightarrow \Lambda_{\rm eff}$, the DCN instantons will describe the
nucleation of a dilatonic black hole pair immersed in an inflating
universe. A detailed discussion of this effect is beyond the scope of
this paper and will be presented separately. We shall only outline
some of the expected results below.

In the absence of any rapidly evolving variables, one can approximate
the background action by the de~Sitter value~\cite{BouHaw95}:
\begin{equation}
I_{\rm DdS} = - \frac{3\pi}{2 \tilde{\Lambda}_{\rm eff}}.
\end{equation}
Thus we obtain for the pair creation rate during inflation:
\begin{equation}
\Gamma_{\rm DCN} =
\exp \left[ - 2 \left( I_{\rm DCN} - I_{\rm DdS} \right) \right]
  = \exp \left[ - \frac{\pi}{\tilde{\Lambda}_{\rm eff}} \:
    \frac{1+3|\tilde{g}|}{1+|\tilde{g}|} \right].
  \label{eq-pcr-DCN}
\end{equation}
Again the form is similar to the result without a dilaton,
Eq.~(\ref{eq-pcr-CN}). But the fact that ${\tilde{g}}$ is not a
degree of freedom will lead to an interesting qualitative difference.
In inflation without a dilaton, the value of the inflaton field fixes
the background effective cosmological constant, but $g$ remains a
degree of freedom.  Therefore black holes can be produced at any time
during inflation, as long as $\Lambda_{\rm eff}$ is small enough to
allow at least one unit of charge~\cite{BouHaw96}. In dilatonic
inflation, however, both degrees of freedom, ${\tilde{\Lambda}}_{\rm
  eff}$ and $Q$, are fixed by the values of the inflaton and dilaton
fields. Since we would still expect the charge to be quantised, this
means that black holes cannot be produced continuously throughout
inflation, but only when the inflaton and dilaton fields have values
that correspond to integer $Q$.

\sect{Charged Nariai Solutions With a General Dilaton Potential}
\label{sec-general}

In Sec.~\ref{sec-DCN} we considered a dilatonic potential of the
Liouville type, for which the DCN solutions are particularly easy to
analyse. Now we shall find the conditions for more general potentials
to admit DCN solutions. As an example we will analyse the case of a
massive dilaton.

Consider now a Lagrangian with a general dilaton potential $V(\phi)$,
\begin{eqnarray}
L = (-g)^{1/2} \left( R - 2(\nabla \phi)^2 - 2 V(\phi)
    - e^{-2a\phi} F^2 \right).
  \label{eq-gen-dil-lagrangian}
\end{eqnarray}
(In Secs.~\ref{sec-DCN} and \ref{sec-dil-euc} we chose $V(\phi) =
\Lambda e^{2b\phi}$.) The following definitions generalise the
quantities $ \tilde{\Lambda} $, $b$ and $ \tilde{g} $ that appeared in
the previous sections:
\begin{eqnarray}
\tilde{\Lambda}(\phi) & = & V(\phi), \\
b(\phi)               & = & \frac{V'}{2V}, \\
\tilde{g}(\phi)       & = & \frac{b(\phi)}{a}, \\
\end{eqnarray}
where a prime denotes differentiation with respect to $\phi$. With
these definitions, DCN solutions (if they exist) will again be given
by Eqs.~(\ref{eq-dilCN-metric}), (\ref{eq-tildes-to-AB}),
(\ref{eq-dil-F-mag}) and (\ref{eq-dil-F-el}).  For constant $\phi$,
the dilaton equation of motion becomes
\begin{equation}
a\, e^{-2a\phi}\, F^2 = V',
\label{eq-dil-3}
\end{equation}
or equivalently,
\begin{equation}
  \tilde{g}\, e^{2a\phi} = (1+|\tilde{g}|)^2\, Q^2
  \tilde{\Lambda}.
\label{eq-dil-4}
\end{equation}
Note that both $ \tilde{\Lambda} $ and $ \tilde{g} $ depend on $\phi$.
Generalising from Sec.~\ref{sec-DCN}, we note that DCN black holes
exist if and only if this equation has at least one solution $\phi_0$
with $V(\phi_0) >0$ and $|\tilde{g}(\phi_0)|<1$. The first condition
ensures the positivity of the cosmological constant, while the second
ensures the positivity of the metric coefficient $\tilde{A}$. Then it
follows from Eq.~(\ref{eq-dil-4}) that the black holes are magnetic
(electric) for positive (negative) $\tilde{g}$. For $\tilde{g} =0$ one
obtains the non-dilatonic neutral Nariai solutions.

Let us consider a massive dilaton potential as an example:
\begin{equation}
V_{\rm massive} = m^2 \phi^2.
\end{equation}
Then the Lagrangian is invariant under the transformation
\begin{equation}
  a \rightarrow -a,\ \phi \rightarrow -\phi.
\end{equation}
Again we fix a gauge by choosing $a \geq 0$. For this potential,
Eq.~(\ref{eq-dil-4}) becomes
\begin{equation}
e^{-2a\phi}\, \phi\, (a\phi+1)^2 = \frac{a}{Q^2 m^2}.
\label{eq-dil-5}
\end{equation}
The solution $\phi=0$, $a=0$ is excluded because of the condition
$V(\phi_0) >0$. $\phi$ must be of the same sign as $a$; thus, with our
gauge choice, $\phi>0$. We cannot solve Eq.~(\ref{eq-dil-5}) in closed
form, but the constant value of $\phi$ is of little interest anyway.
Note, however, that the left hand side of Eq.~(\ref{eq-dil-5}) is zero
for $\phi=0$, then increases to reach a maximum at $\phi=1$, and then
decreases, approaching zero as $\phi \rightarrow \infty$. The value of
the maximum is $4/ae^2$, where $e$ is Euler's number. Thus the charge
$Q$ has a {\em lower} bound,
\begin{equation}
|Q| \geq \frac{ea}{2m},
\end{equation}
but no upper bound.

\sect{Summary}

Charged Nariai solutions are important because they have a regular
Euclidean section and can therefore mediate pair creation processes.
Unlike all other RNdS black holes, the Charged Nariai solutions of
Einstein-Maxwell theory possess an analogue in dilatonic theories with
a Liouville potential. The metric and Maxwell field of the DCN
solutions can be written in a form very similar to the CN solutions.
However, there are parameter pairs which parametrise the CN solutions,
but whose dilatonic analogues do not parametrise the DCN solutions.
Consequently, the solution spaces differ significantly. We found that
the shape of the geometry is entirely fixed by the ratio of the
dilaton coupling constants, $\tilde{g} = b/a$. The black hole charge
only affects the overall scale of the metric. The charge can thus be
arbitrarily large; by Eq.~(\ref{eq-dil-2}), however, a large charge
will lead to a large value of the dilaton, and thus to a very small
Maxwell coupling. As a consequence, the force on a charged test
particle turns out to be independent of the black hole charge. The
electro-magnetic coupling strength can be measured using two test
particles; this in turn determines the black hole charge. The DCN
solution space is two-dimensional only for $0<|\tilde{g}|<1$. If $
\tilde{g} =0$, only the non-dilatonic neutral Nariai solution can be
found. For other values of $\tilde{g}$, no DCN solutions are admitted.
The sign of $b$ determines whether the black hole is magnetic,
neutral, or electric. 

The Euclidean DCN solutions themselves do not describe a pair creation
process, since there is no dilatonic de~Sitter solution which could
act as a background. If the fixed cosmological constant is replaced by
a slowly decreasing effective cosmological constant, however, one
would expect the DCN instanton to mediate black hole pair creation
during extended chaotic inflation.

We found conditions for the existence of Charged Nariai black holes in
theories with a general dilaton potential. For the case of a massive
dilaton, we showed that such solutions exist if the black hole is
sufficiently charged.

\section*{Acknowledgements}

I am grateful to Stephen Hawking for suggestions which led to this
work.  I wish to thank him as well as Gary Gibbons and Simon Ross for
their valuable comments on a draft of this paper, and for many
interesting discussions. I gratefully acknowledge financial support
from EPSRC, St John's College and the Studienstiftung des deutschen
Volkes.

\end{document}